\documentclass[12pt]{article}
\usepackage{graphicx}
\newcommand{\PSbox}[3]{\mbox{\rule{0in}{#3}\includegraphics{#1}\hspace{#2}}}
\newcommand{\beq}{\begin{eqnarray}}
\newcommand{\eeq}{\end{eqnarray}}
\def\be{\begin{equation}}
\def\ee{\end{equation}}
\def\ba{\begin{eqnarray}}
\def\ea{\end{eqnarray}}

\newcommand{ \sla }[1]{\setbox0=\hbox{$#1$}         
   \dimen0=\wd0                                     
   \setbox1=\hbox{/} \dimen1=\wd1                   
   \ifdim\dimen0>\dimen1                            
      \rlap{\hbox to \dimen0{\hfil/\hfil}}          
      #1                                            
   \else                                            
      \rlap{\hbox to \dimen1{\hfil$#1$\hfil}}       
      /                                             
   \fi}                                             %

\def\lesssim{\mathrel{\mathpalette\vereq<}}

\makeatletter
\def\vereq#1#2{\lower3pt\vbox{\baselineskip1.5pt \lineskip1.5pt
\ialign{$\m@th#1\hfill##\hfil$\crcr#2\crcr\sim\crcr}}}
\makeatother

\begin{document}

\begin{titlepage}
\noindent
\begin{flushright}
SU-ITP-01/55\\
{\tt hep-ph/0112212}\\
\end{flushright}

\vskip1.5cm
\begin{center}
{\Large {\bf Effects of the Intergalactic Plasma on Supernova}} \\
\vskip .05in
{\Large {\bf Dimming via Photon-Axion Oscillations}} \\

\end{center}
\vskip1cm

\begin{center}
{\bf Csaba Cs\'aki$^{a,b}$,
Nemanja Kaloper$^{c}$ and
John Terning$^{a}$}
\end{center}

\vskip 10pt

\begin{center}
$^a${\em Theoretical Division T-8, Los Alamos National Laboratory,
Los Alamos, NM 87545}\\

\vskip 0.1in

$^b${\em Newman Laboratory of Physics, Cornell University, Ithaca, NY 14853}\\

\vskip 0.1in

$^c${\em Department of Physics, Stanford University,
Stanford CA 94305-4060 }\\

\vskip 0.1in

\vskip 0.1in
{\tt  csaki@mail.lns.cornell.edu, kaloper@stanford.edu, terning@lanl.gov}

\end{center}

\begin{abstract}
We have recently proposed a mechanism of photon-axion
oscillations as a way of rendering supernovae dimmer without
cosmic acceleration. Subsequently, it has been argued that 
the intergalactic plasma may
interfere adversely with this mechanism by rendering the oscillations
energy dependent. Here we show that this energy
dependence is extremely sensitive to the precise value of the free
electron density in the Universe. Decreasing the
electron density by only a factor of 4 is already sufficient to bring 
the energy dependence within the experimental bounds.
Models of the intergalactic medium show that
for redshifts $z<1$ about 97\% of the total volume of
space is filled with regions of density significantly 
lower than the average density. From these models we estimate that the 
average electron density in most of space 
is lower by at least a factor of 15 compared to the estimate
based on one half of all baryons being uniformly distributed and ionized. 
Therefore the energy
dependence of the photon-axion oscillations is consistent with
experiment,
and the oscillation model remains a viable alternative to the accelerating
Universe for explaining the supernova observations.
Furthermore, the electron density does give rise to a sufficiently large 
plasma
frequency which cuts off the photon-axion mixing above microwave frequencies,
shielding the cosmic microwave photons from axion conversions and significantly
relaxing the lower bounds on the axion mass implied by the oscillation model.
\end{abstract}
\end{titlepage}
\vskip0.5cm

\setcounter{equation}{0} \setcounter{footnote}{0}

We recently proposed~\cite{us} that photon-axion
oscillations \cite{rafsto} in external magnetic fields, 
could  explain the observed~\cite{supnov}
dimmer supernovae  at redshifts
$0.3 \leq z\leq 1.7$. This mechanism requires the photon-axion coupling
scale to be $M\sim 4 \cdot 10^{11}$ GeV and assumes an 
axion mass $m\sim 10^{-16}$ eV, 
and an intergalactic magnetic field $B\sim 5 \cdot
10^{-9}$ G, with a
domain size of order a Mpc, in agreement with observational bounds
\cite{kronberg}. Detailed properties of the model were
investigated in~\cite{jecg}.
The oscillation mechanism rests on two key ingredients:

\begin{itemize}
\item
The photon-axion oscillation probability of photons with energy ${\cal E}$
induced by the off-diagonal element $|{\cal M}_{12}|={\cal E} B/M={\cal E}\mu$
of the mixing matrix ${\cal M}$ \cite{us} should be non-negligible
over a significant fraction of
the line of sight, e.g. over a few thousand magnetic domains;
\item
The magnetic field of essentially random direction
over many domains will induce a decrease of photon flux saturating at
about 2/3 of its original value,
independently of the detailed structure of the magnetic field
and the precise form of the mixing,
provided the mixing is not negligibly small.
\end{itemize}

\begin{figure}
\PSbox{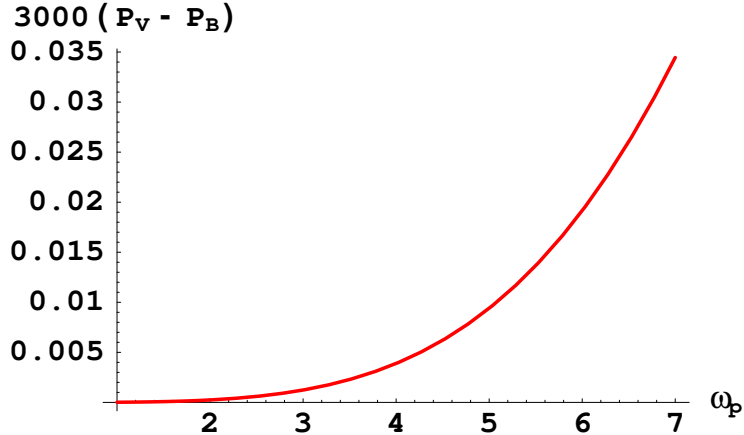 hscale=100 vscale=100 hoffset=0
voffset=-10}{7cm}{6cm}
\begin{center}
\parbox{5in}{\caption{The difference in photon survival probability
difference for two
energy bands summed over 3000 domains of size 1 Mpc. The horizontal axis is
the photon plasma frequency in units of $10^{-15}$ eV.
\label{fig:difference}}}
\end{center}
\end{figure}

Subsequently the authors of~\cite{dhuz} argued that photon rescattering
in the ionized gas comprising the intergalactic medium (IGM) could have
rendered the oscillations very energy-dependent, which would have been
inconsistent with observations.
Namely in the presence of ionized electrons
the mixing matrix is
\begin{equation}
{\cal M} = \pmatrix{ \omega_p^2 & i {\cal E} \mu \cr
-i {\cal E} \mu & m^2 \cr} \, ,
\label{mixing}
\end{equation}
where $\omega_p$ is the plasma frequency \cite{rafsto}
$\omega_p^2=4\pi \alpha n_e/m_e$, with $n_e$ the electron density,
$m_e$ the electron mass, and $\alpha$ the fine structure constant.
Then the conversion probability
of photons into axions over a domain of size
$L_{dom}$ is given by
\begin{equation}
P_{\gamma \rightarrow a} = \frac{4 \mu^2 {\cal E}^2}{(\omega_p^2-m^2)^2+4\mu^2
{\cal E}^2}
\sin^2\left[  \frac{\sqrt{(\omega_p^2-m^2)^2 + 4 \mu^2 {\cal E}^2}}{4{\cal E}}
L_{dom}\right]
\, .
\label{prob}
\end{equation}
Using this formula one can estimate the
largest allowed value of $\omega_p^2$ for which the energy dependence
of the dimming effect is within the observational bounds.
Here we assume that the magnetic domain size is
$\sim 1$ Mpc~\cite{kronberg}, and that there are about 3000 domains
for supernovae at redshifts $z \lesssim 1$. We consider photons
of average energy of 4.3 eV in the B band and average energy
of 3.4 eV in the V band, and require that the total difference
in oscillation probabilities, i.e. for photon survival as a function
of energy, is less than 3 percent~\cite{supnov}.
These energy values are slightly higher
to reflect the redshift of the photon energies as they journey
from $z=1$ to $z=0$; we simply pick the energies at $z=0.5$.
The resulting difference in the oscillation probabilities
is displayed in Fig.~\ref{fig:difference}.
This figure is probably an overestimate of the energy dependence since we
have simply added the probabilities over all the domains, while in reality
the photon conversion saturates at a value of $1/3$
due to the random direction of the magnetic field
in different domains. Even so, we can see
that if the plasma frequency obeys  $\omega_p \leq 6\cdot 10^{-15}$ eV
the energy dependence disappears very rapidly, and becomes
undetectable by present observations.
This threshold value of $\omega_p$
corresponds to a free electron density of $n_e=2.5 \cdot
10^{-8} {\rm cm}^{-3}$.

The authors of \cite{dhuz} chose the value of the electron density
to be $n_e=1 \cdot
10^{-7} {\rm cm}^{-3}$. This is barely above the threshold
below which the energy dependence of oscillations becomes
unobservable. On the other hand,
if all of the baryons in our Universe had been uniformly distributed
and fully ionized, the electron density would have
been\footnote{Assuming that $\Omega_B \simeq 0.045$ is the baryon fraction of 
the
energy density of the Universe, the Hubble parameter is $H=65$ km/s/Mpc,
the critical density is $\rho_c \simeq 4.3 \cdot
10^3 ~{\rm eV} {\rm cm}^{-3}$, and
that 10\% of baryons are neutrons in helium nuclei.
We are ignoring the tiny fractions of heavier elements.}
$n_e=1.8 \cdot 10^{-7} {\rm cm}^{-3}$.
The value assumed
in \cite{dhuz} corresponds to roughly
a half of all baryons being
uniformly distributed and ionized. This estimate
is very likely too high, by at least a factor of 15,
in the region $z \lesssim 1$ which
is the relevant regime for our model
since most of the observed supernovae reside in this range.
This reduction of $n_e$
would render the energy dependence unobservable at present.

Let us now discuss why the electron
density has to be reduced. It is well known that the distribution of baryons
at low redshifts $z \sim 0$ is
an open problem in cosmology \cite{Fukugita}. 
At low redshifts the baryons are
thought to be divided among the
following structures in the Universe \cite{Fukugita,Peebles,Dave,Cen}:

\vskip.3cm
\indent $\bullet$ Condensed baryons in stars and galactic gas

\indent $\bullet$ Hot baryons in galaxy clusters and groups

\indent $\bullet$ WHIM: warm-hot intergalactic matter

\indent $\bullet$ photoionized intergalactic gas (Lyman $\alpha$ forests).
\vskip.3cm

\noindent At low redshifts the baryons are roughly equally distributed between
the condensed and hot baryons, the WHIM, and the Lyman $\alpha$
forests. The former structures are very strongly clumped,
occupying a tiny fraction of total volume of space at $z \lesssim
1$. Therefore they
are unimportant for the photon-axion oscillations.

Thus the only relevant baryons are about $1/3$ of the total, residing in the
Lyman $\alpha$ forests. They comprise a diffuse photoionized
gas with relatively low temperature, $T< 10^5$ K.
But even if all of this gas
were uniformly distributed, the
electron density would have been
at most $6.1 \cdot 10^{-8} {\rm cm}^{-3}$.
However, the low
temperature and the highly ionized nature of this gas
makes the low density regions extremely difficult to observe
directly. There are no direct
observational bounds on the distribution of low density
Lyman $\alpha$ forests, and hence no direct evidence for the
distribution at $z \lesssim 1$ to be uniform.
Instead, most of the standard lore concerning low-density, low-redshift
plasma comes from model-dependent simulations.

While the assumption for uniformity would be very accurate
for large values of redshifts ($z>4-5$), for low $z$ even the
low-density gas in the Lyman $\alpha$ forests tends to clump.
The simulations and the existing observations about the
regions of gas with significantly higher than average density
show that it is unlikely that gas in the Lyman $\alpha$ forests would
be uniformly distributed in space.
Indeed, at $z\sim 0$ the average overdensity of
these kinds of structures
is in the range 10--100 \cite{lowredshift}.
Their characteristic
radial size is $\lesssim 100$ kpc \cite{shaye}, and they
take up only a small fraction of space.
Almost all the  observed Lyman $\alpha$ forests fall into
this category. These forests are  correlated with galaxies: they tend
to reside around the halos of galaxies. The
simulations also display a correlation of even the
low density forests with the large scale structure of the Universe,
supporting the claim that for low redshifts even this gas is clumped.
In the analytic model of ref. \cite{silk} at low redshifts
about 97 \% of space is filled with low-density gas, which is underdense by at least
a factor of 10. Therefore it is
reasonable to assume that over most of space
at redshifts $z \lesssim 1$ the electron density is at most
$n_e \leq 6 \cdot 10^{-9} {\rm cm}^{-3}$, and probably even less than
that. In the absence of direct observational evidence
one cannot of course completely exclude somewhat
larger densities, but all the indirect evidence favors
the above number as a reasonable bound at low redshifts.
With this value of the electron
density the plasma frequency is $\omega_p \leq 3\cdot 10^{-15}$
eV. This places it safely in the regime
where the energy dependence of the supernova dimming is
below the current experimental sensitivity.

A closer scrutiny actually reveals that
the effects of intergalactic plasma are beneficial for our model because
they make it insensitive to the axion mass as long as it is below the
plasma frequency.
Previously, we had selected the value of the
axion mass to suppress the mixing between 
cosmic microwave background (CMB) photons 
and axions, giving
$m \sim {\rm few} \cdot \, 10^{-16}$ eV. However, the plasma-generated 
effective 
photon mass
has the right magnitude to suppress the photon-axion mixing at an energy above
the CMB photon energies.
This
relaxes the lower bound on the axion mass for the model, since now the
axion mass could be considerably smaller\footnote{We refrain from
attempting to formulate the precise bounds on the axion mass
since it is believed that
at some value of redshift $z > 6$ the IGM was neutral, removing the
plasma and therefore the photon mass. At present it is not known how
the background magnetic fields emerged, and so they could still have
been large enough at such high  redshifts to affect CMBR if the axion
was too light.}.

\begin{figure}
\PSbox{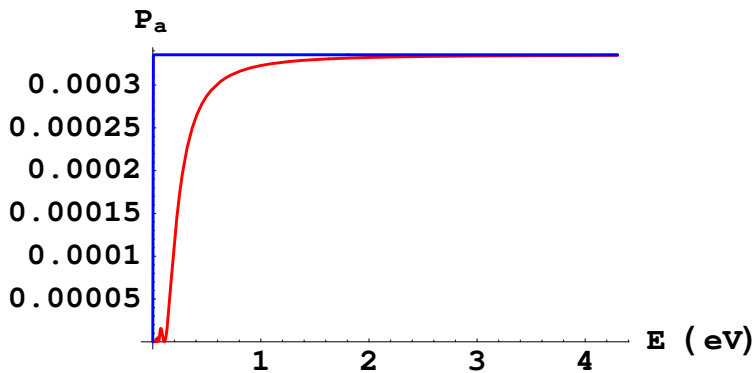 hscale=100 vscale=100 hoffset=0
voffset=-10}{7cm}{6cm}
\begin{center}
\parbox{5in}{\caption{The photon oscillation probability for
one domain of size 1 Mpc versus photon energy. The top curve is for the
model with an axion mass of $10^{-16}$ eV and no plasma, while the bottom
curve is for a photon plasma frequency of $ 3 \cdot 10^{-15}$ eV.
\label{fig:energy}}}
\end{center}
\end{figure}

One may also ask if the axion-photon coupling should be
changed because of the plasma-induced photon mass. In
Fig.~\ref{fig:energy} we have plotted the oscillation probability in a magnetic
domain as a function of energy for the original scenario with
$\omega_p = 0$, and for the case with $\omega_p=3\cdot 10^{-15}$ eV. One sees
immediately that for optical photons, ${\cal E}>2$ eV, the change of the oscillation
probability in a domain is very small, with the main difference being that the
oscillations are cut off at energies closer to the optical
regime than before. Thus the plasma effects do not require a
significant alteration of the axion-photon coupling.

In summary, we have considered the effect of the IGM
on our proposed mechanism \cite{us} for supernova dimming via photon-axion
oscillations. We have found that the effects of the intergalactic 
plasma are very
sensitive to the precise value of the plasma frequency.
We have put a conservative bound on the plasma frequency by combining the
observations and the simulations of the IGM, and found that the plasma 
effects
would not strongly influence the oscillations of optical photons. Instead
they would provide a natural cutoff on the oscillations at energies below the
optical range; this relaxes the lower bounds on the mass of the 
axion.
We conclude that the photon-axion oscillation mechanism
remains a viable alternative to the accelerating Universe for
explaining the supernova observations.

\section*{Acknowledgements}
We thank Scott Armel-Funkhouser,
Christophe Grojean, Georg Raffelt and Martin White
for useful discussions and correspondence.
C.C. is an Oppenheimer fellow at the Los Alamos National Laboratory, and is
supported in part by a DOE OJI grant.
C.C. and J.T. are supported by the U.S. Department
of Energy under contract W-7405-ENG-36. N.K. is supported in part
by an NSF grant PHY-9870115.

\end{document}